\documentclass[review,times,a4paper]{elsarticle}

\usepackage[USenglish]{babel}
\usepackage[utf8]{inputenc}
\usepackage[T1]{fontenc}

\usepackage{natbib}
\setcitestyle{authoryear}
\usepackage{amsmath}
\usepackage{subcaption}
\usepackage{booktabs}
\usepackage{siunitx}
\usepackage{amssymb}
\usepackage{graphicx}
\usepackage{caption}
\usepackage{multirow}
\usepackage{color,soul}
\usepackage[super]{nth}
\usepackage{graphicx}
\usepackage[ruled,vlined]{algorithm2e}
\usepackage{float}
\usepackage{url}
\usepackage{ulem}
\usepackage[noend]{algpseudocode}
\usepackage{hyperref}
\usepackage[capitalize]{cleveref}
\usepackage[version=4]{mhchem}
\usepackage{scrextend}
\graphicspath{{./figures/}}

\def\stress{\boldsymbol{\sigma}}

\def\strain{\boldsymbol{\varepsilon}}
\def\eigen{\strain_{eig}}

\def\coord{\boldsymbol{x}}

\def\eshin{\mathbb{S}}
\def\eshext{\mathbb{G}(\coord)}
\def\iden{\mathbb{I}}
\def\stif{\mathbb{C}^0}

\def\sif1{K_I}
\def\tough{K_{IC}}
\def\wf{{h}(l,y')}
\def\fren{G_{C}}

\begin{document}

\begin{frontmatter}
  \title{Predicting damage in aggregates due to the volume increase of
    the alkali-silica reaction products}
  
  \author[epfl]{E.R.~Gallyamov\corref{cor1}}\ead{emil.gallyamov@epfl.ch}
  \author[empa]{A.~Leemann}
  \author[empa,ntnu]{B.~Lothenbach}
  \author[epfl]{J.-F.~Molinari}
  \cortext[cor1]{Corresponding author}
  
  \address[epfl]{Civil Engineering Institute, Materials Science and
    Engineering Institute, École Polytechnique Fédérale de Lausanne
    (EPFL), Station 18, CH-1015 Lausanne, Switzerland}
  \address[empa]{Laboratory for Concrete \& Construction Chemistry, Swiss
    Federal Laboratories for Materials Science and Technology (Empa),
    8600 Dübendorf, Switzerland}
  \address[ntnu]{Department of Structural
    Engineering, Norwegian University of Science and Technology (NTNU),
    7491 Trondheim, Norway}
  
  \begin{abstract}
    Volume increase between the reactants and the products of alkali
    silica reaction could reach up to $100\%$. Taking place inside the
    aggregates, ASR imposes internal pressure on the surrounding
    material. In the current paper, we study the possibility of crack
    growth due to such internal loading. This study is done by employing
    a semi-analytical mechanical model comprising an elastic solution to
    a well-known Eshelby problem and a linear elastic fracture mechanics
    solution to a ring-shaped crack encircling a spheroidal
    inclusion. The proposed method implies the availability of
    pre-existing micro-fissures within the aggregate.

    The study reveals dependence of the crack growing potential on the
    spheroid's shape: the larger is the ASR pocket - the longer crack it
    can open. Two most critical shapes, causing a highest stress
    intensity factor and developing the longest crack, are a sphere and a
    spheroid with $1/4$ aspect ratio respectively. The size analysis of
    the problem suggests a critical spheroid's radius below which no
    crack growth is expected. For a chosen material properties and
    expansion value, such radius lays in the range between $0.1 \ \mu m$
    and $1 \ \mu m$. Independently of the expansion value and the shape
    of the pocket of ASR product, the maximum crack length has a
    power-law dependence on the size of a spheroid.

    All the theoretical predictions are confirmed by a numerical model
    based on the combination of the finite element method and cohesive
    element approach.
  \end{abstract}
  
\end{frontmatter}

\section{Introduction}

Alkali-silica reaction (ASR) in concrete is the reaction between
\ce{SiO2} contained in aggregates and alkalies coming from the cement
paste \citep{swamy_alkali-silica_2003}. ASR products accumulate within
the aggregate leading to build-up of stress, expansion of concrete and
evolution of cracks. ASR causes substantial damage to the concrete
infrastructure worldwide.

The size of the initial ASR products, before aggregates cracks, is in the
range of $10$s of nanometres to a few micrometres. They form between
adjacent mineral grains within reactive aggregates close to the border
with the cement paste
\citep{leemann_types_2016,leemann_addition_2019}. Shapes of the expanding
products are highly irregular as a result of aggregate porosity and
pre-existing defects that are common in aggregates. This primary ASR
product yields internal loading on aggregates leading to
cracking. Resulting cracks originate in the aggregates and extend into
cement paste. Opening of cracks could be much larger than the initial ASR
product size and pre-existing cracks. As reaction advances, the alkali
front moves inwards the aggregates, producing more and more ASR
sites. The newly formed cracks start to fill with the secondary ASR
products.

The most widely reported hypothesis for the expansion of concrete caused
by ASR is swelling of the ASR products due to the absorbance of
water. Recent water uptake measurements on synthetic ASR products,
however, indicate a very limited water uptake
\citep{shi_synthesis_2019}. This suggests that ASR expansion is not
caused by swelling of the ASR products but by alternative
mechanisms. Even if the underlying mechanism for expansion is not known,
it is clear that the formation of ASR products leads to an increase in
molar volume. ASR products, commonly reported in the literature,
precipitate as an amorphous gel. However, it has been shown that in
addition to amorphous ASR product, the one present in large cracks within
aggregates could be crystalline \citep{cole_products_1983,
  de_ceukelaire_determination_1991,dahn_application_2016,
  leemann_characterization_2020}.

The goal of this study is to verify the hypothesis if the gel-like ASR
product could initiate cracks due to expansion. If this is the case, the
role of the shape and size of the pocket of ASR product in this process
is to be investigated. First, the responsible chemical reaction is
identified and the increase in the volume of the products over the
reactants is computed. Then, a semi-analytical mechanical model for an
expanding ASR pocket surrounded by a pre-existing micro-crack is
formulated. Crack extension is predicted for the ASR pockets of
different sizes, shapes and expansion values. Finally, the predictions
are verified by a numerical finite element model.

\section{Methods}
\label{sec:methods}

\subsection{Chemical reaction}
\label{sec:chemical-reaction}

A typical ASR formation reaction could be written as
\begin{equation}
  \ce{ $\underset{\text{solid}}{\ce{4SiO2}}$ + K+ + Ca^2+ + 3OH- + 2H2O -> $\underset{\text{ASR product}}{\ce{KCa[Si4O9(OH)]*3H2O}}$},
\end{equation}
which describes the formation of an ASR product with molar ratios K/Si
and Ca/Si of $0.25$. These ratios correspond to the elemental ratios
observed in ASR products in field samples as well as to those observed in
laboratory synthesised ASR products
\citep{leemann_e-modulus_2013,shi_synthesis_2019,shi_formation_2020}.

\begin{sloppypar}
  Assuming quartz or amorphous \ce{SiO2} as silica source, it is possible
  to calculate the increase of solid volume if ASR products are formed
  within the space originally occupied by \ce{SiO2}. Note that other
  species (\ce{K+}, \ce{Ca^2+}, \ce{OH-}, and \ce{H2O}), needed to form
  ASR products, are assumed to diffuse into the aggregate from the cement
  pore solution, such that their volume is not considered. Based on the
  molar volume of quartz ($22.6$ cm$^3$/mol), and the volume of
  crystalline ASR product K-shlykovite, \ce{KCa[Si4O9(OH)] * 3H2O}, of
  $183$ cm$^3$/mol \citep{geng_mechanical_2020} (or $45.7$ cm$^3$/mol if
  normalised to $1$ silica: \ce{K_{0.25}Ca_{0.25}[SiO_{2.25}(OH)_{0.25}]
    * 0.75H2O}) an increase in solid volume by a factor of two is
  obtained, and a ratio between the additional volume
  ($45.7-22.6 = 23.1$) and the initial volume is $23.1/22.6 \approx 1.0$,
  thus leading to $100\%$ expansion. For amorphous ASR products, the
  molar volume has not been measured, but based on the extensive analysis
  performed in \citep{leemann_characterization_2020}, amorphous ASR
  product with a composition of
  \ce{Na_{0.1}K2_{0.2}Ca2_{0.2}SiO_{2.3}(OH)_{0.1} * 1.1H2O} indicates a
  similar or even higher volume increase ($\approx 2.3$) as for
  crystalline ASR product. The duplication of the initial volume due to
  the precipitation of ASR product could potentially lead to formation of
  cracks within aggregates. This hypothesis was studied in details based
  on the semi-analytical mechanical model of an inclusion expanding
  within an infinite medium.
\end{sloppypar}

The generation of crystallisation pressures due to over-saturation of the
surrounding solution during the formation of ASR cannot be excluded as a
possible expansion mechanism. This direction is not further investigated
in the present paper as no solution measurements are available, making a
reliable estimation of maximum crystallisation pressures presently not
possible.
 
\subsection{Analytical model of expanding ASR pockets}
\label{sec:analytical}

Dependence of the possible crack growth on the size and shape of ASR
products is studied by exploring a semi-analytical mechanical model,
which consists of two components. The first one is a purely analytical
model for an inclusion expanding within an infinite medium. The second
one is a semi-analytical model of a ring-shaped crack encircling the
previously described inclusion. The final model yields stress
concentrations at the external crack tip that are used to evaluate the
crack growth. A pocket of ASR product is assumed to have an ellipsoidal
shape since the latter comprises a wide range of shapes with central
symmetry (e.g. spheres, oblate spheroids, disks, etc).

The departing point of the analytical model is the Eshelby problem
\citep{eshelby_elastic_1959}, which describes an expanding ellipsoidal
inclusion embedded in an infinite medium. The term ``inclusion'' implies
that its elastic properties differ from the ones of the matrix. The
inclusion represents a single pocket of ASR product surrounded by the
aggregate. Here we assume a relatively small ASR pockets positioned
sufficiently far from the aggregate's boundary not to cause mechanical
influence, which justifies the ``infinite surrounding medium''
condition. Given the known expansion of the inclusion, mechanical
properties of the ASR product and the aggregate, their shapes and sizes,
the stress and strain fields could be determined. The problem was solved
by \cite{eshelby_elastic_1959,eshelby_determination_1957} for general
eigen strains. The solutions for the inclusion and the matrix are
different. Later \cite{mura_micromechanics_1987} provided explicit
equations for the Eshelby tensor $\eshin$ for different shapes of an
ellipsoid at the interior points. \cite{ju_novel_1999,ju_effective_2001}
gave the first explicit formulas for computing the Eshelby tensor in the
surrounding matrix. \cite{healy_elastic_2009} developed a
MATLAB\textsuperscript{TM} code with the Eshelby solution in 3D both
inside and outside the inclusion.
\begin{figure}
  \centering \includegraphics[width=0.5\textwidth]{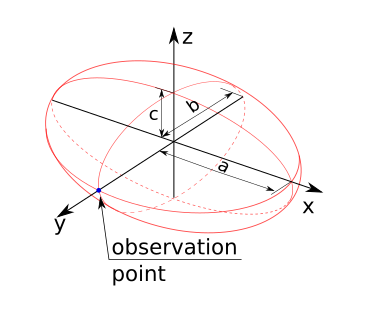}
  \caption{Geometry of an ellipsoid. Further in the text, stresses are
    evaluated either in the observation point or along the $y$-axis}
  \label{fig:ellipsoid}
\end{figure}

The elliptic inclusion with semi-axes $a$, $b$ and $c$ (shown in
\cref{fig:ellipsoid}) is embedded in a matrix. In this study, we have
considered a spheroidal shape with the equatorial radius
$b=c=r$. Variation of the semi-axis $a$ allows to consider different
geometries: 1) sphere ($a = b = c$); 2) needle ($a >> b = c$); 3) penny
($a << b = c$). The expansion is applied at the inclusion in the form of
eigen strain $\eigen$ - strain causing zero stress under no
confinement. The eigenstrain is linked to the elastic strain
$\boldsymbol{\varepsilon}_{el}$ as
\begin{equation}
  \boldsymbol{\varepsilon} = \boldsymbol{\varepsilon}_{el} + \boldsymbol{\varepsilon}_{eig},
\end{equation}
where $\boldsymbol{\varepsilon}$ is the total strain. The matrix is an
infinite body with remote stress $\stress^0$ or strain $\strain^0$
applied. In the current study, we assume stress-free body with
$\stress^0$ and $\strain^0$ equal to zero. The effect of external loading
on ASR was previously studied by multiple authors
\citep[e.g.][]{larive_apports_1997,multon_effect_2006,dunant_effects_2012-1}.

Strain and stress inside the inclusion are given by
\citep{ju_effective_2001}
\begin{equation}
  \begin{split} 
    \label{eq:str_inside}
    &\strain=\strain^0 + \eshin : \eigen,\\
    &\stress=\stress^0 + \stif [\eshin - \iden] : \eigen,
  \end{split}
\end{equation}
where $\eshin$ is the 4th-order Eshelby tensor for interior points,
$\stif$ the matrix stiffness tensor and $\iden$ the identity tensor. The
particular property of the Eshelby's problem is that the stress and
strain fields at the interior points are uniform and the Eshelby tensor
$\eshin$ is independent of the position within the inclusion.
 
Oppositely, for the external field, strain and stress at a point depend
on its position:
\begin{equation}
  \begin{split} 
    \label{eq:str_outside}
    &\strain(\coord)=\strain^0 + \eshext : \eigen,\\
    &\stress(\coord)=\stress^0 + \stif \eshext : \eigen,
  \end{split}
\end{equation}
where $\eshext$ is a 4th-order Eshelby tensor for exterior points.

Explicit formulas for the components of $\eshin$ and $\eshext$, available
in \cite{ju_effective_2001} and \cite{healy_elastic_2009}, yield the
values of $\strain$ and $\stress$ at any point of space both inside and
outside the inclusion.

\subsection{Semi-analytical model with a ring-shaped crack}
\label{sec:semi-analytical}

\citet{griffith_phenomena_1921} in his fundamental work showed that the
material strength at the macro-scale is largely affected by the presence
of the microscopic flaws. For an expanding inclusion, to grow a crack
from its surface, there should be an initial micro-fissure which would
facilitate the crack propagation. Higher expansion would be required if
such an inclusion would be surrounded by sound material with no
pre-existing flaws. Pre-existing defects in the form of microcracks are
common in aggregates.

\begin{figure}
  \centering \includegraphics[width=0.5\textwidth]{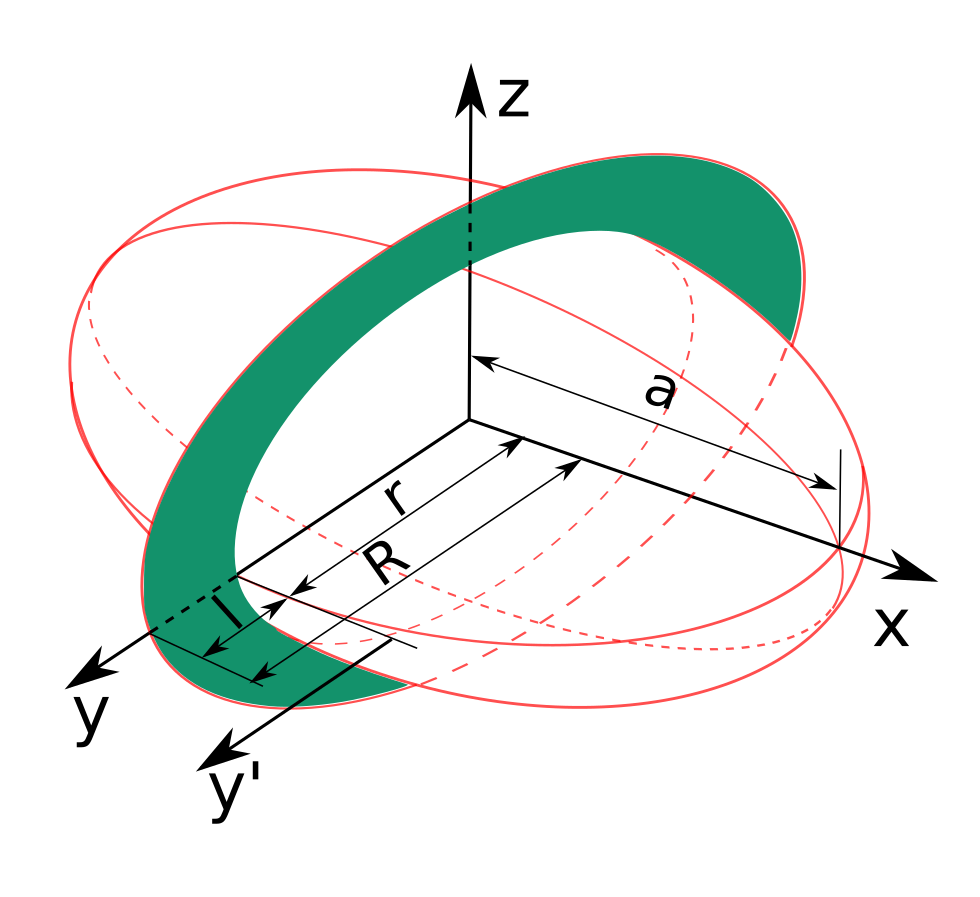}
  \caption{Spheroid with the pre-existing disk-shaped crack in the
    $yz$-plane.}
  \label{fig:ell_crack}
\end{figure}

A crack is added to the Eshelby problem as shown in \cref{fig:ell_crack}.
It has a shape of a flat ring in the $yz$-plane. While the inner crack
tip lays exactly on the surface of the spheroid, the outer one extends to
the distance $l$ from the surface. $r$ and $R$ denote the internal and
external crack radii correspondingly. An additional $y'$-axis was
introduced for illustration purpose: it is a $y$-axis with the origin
shifted to the surface of the spheroid. The crack is opening in mode I
under the loading symmetric with respect to the crack plane.
Keeping the expansion value fixed, different alignment of a crack would
reduce the probability of its extension.

According to the adopted model, the inclusion is always expanding as a
solid body. When the crack extends, the ASR product does not flow
inside. Therefore this model represents the ASR product in its mature
stage. Accounting for the gel fluidity would require an enriched
analytical model including mass conservation and pressure redistribution.

To evaluate the crack growth, one has to know $\sif1$ - stress intensity
factor (SIF) in mode I for the external crack tip. The principle of
superposition states that applying $\stress(\coord)$ on the crack faces
is similar to loading the cracked body with loads that cause
$\stress(\coord)$ in the absence of a crack
\citep{petroski_computation_1978}. To find $\sif1$, we replace the
problem of a crack loaded by a spheroidal inclusion by a crack with its
faces loaded by the stresses caused by the inclusion in the absence of
the crack. The latter ones are known from the Eshelby solution.

Literature has few analytical solutions for computing $\sif1$ for
ring-shaped cracks under simple loading
(e.g. \cite{tada_stress_2000}). Unfortunately, solutions for a general
stress distribution are not available. Instead, this problem could be
solved by means of weight functions, as proposed by
\citet{bueckner_novel_1970} and \citet{rice_remarks_1972}. These authors
suggested to use the known crack face displacement $u(l,x)$ and the known
stress intensity factor $\sif1^0$ for a symmetrical load system on a
cracked body to compute unknown $\sif1$ for any other symmetrical load
system. The equation for SIF reads as
\begin{equation}
  \label{eq:sif_w_f}
  \sif1=\int^r_0 \sigma(y') \wf dy',
\end{equation}
where $\wf$ is a weight function defined as
\begin{equation}
  \label{eq:w_f}
  \wf=H \frac{\partial u}{\partial r}/\sif1^0.
\end{equation}
In \cref{eq:sif_w_f,eq:w_f}, $l$ is the crack length equal to the
difference between the external and internal radii $(R - r)$, $H$ is a
material parameter, $\sigma(y')$ is the stress distribution along the
crack plane in the unflawed body under the load associated to
$\sif1$. Integration is carried out along the crack surface
only. \cite{fett_weight_2007} computed weight functions for a ring-shaped
crack by the interpolating procedure. The latter is the reason for
solution being ``semi-analytical''. The weight function for the outer
crack tip reads as
\begin{equation}
  \label{eq:w_f_fett}
  \begin{split}
    \wf\sqrt{l}=\frac{2}{\sqrt{\pi (1 - y'/l)(l/r + 1)}}
    \left(\frac{1 + y'/r}{\sqrt{2 + l/r + y'/r}} - \right.\\
    \left. \frac{1 - \sqrt{y'/l}}{\sqrt{l/r + 2}}
      \vphantom{\int_1^2}\right).
  \end{split}
\end{equation}
Plugging \cref{eq:w_f_fett} and the stresses from \cref{eq:str_outside}
into \cref{eq:sif_w_f} and integrating it numerically results in required
values of $\sif1$. Similar approach was previously used by
\cite{iskhakov_expansion_2019} within a multi-scale model. This solution
is built on the principles of linear elastic fracture mechanics (LEFM).

\subsection{Numerical validation}
\label{sec:num-valid}
\defcitealias{noauthor_akantu_2021}{akantu.ch}
Analytical and semi-analytical predictions are validated through
numerical modelling. Two components of the final solution are
verified. The first component is the elastic stresses in the inclusion's
surroundings in the absence of the crack. The second component is the
actual crack radius for different inclusion shapes and sizes. The
numerical model is based on the finite element method (FEM). All the
simulations are performed using the open-source parallel FE library
Akantu
[\citealt{richart_implementation_2015}, \citetalias{noauthor_akantu_2021}].

\begin{figure}
  \centering \includegraphics[width=0.8\textwidth]{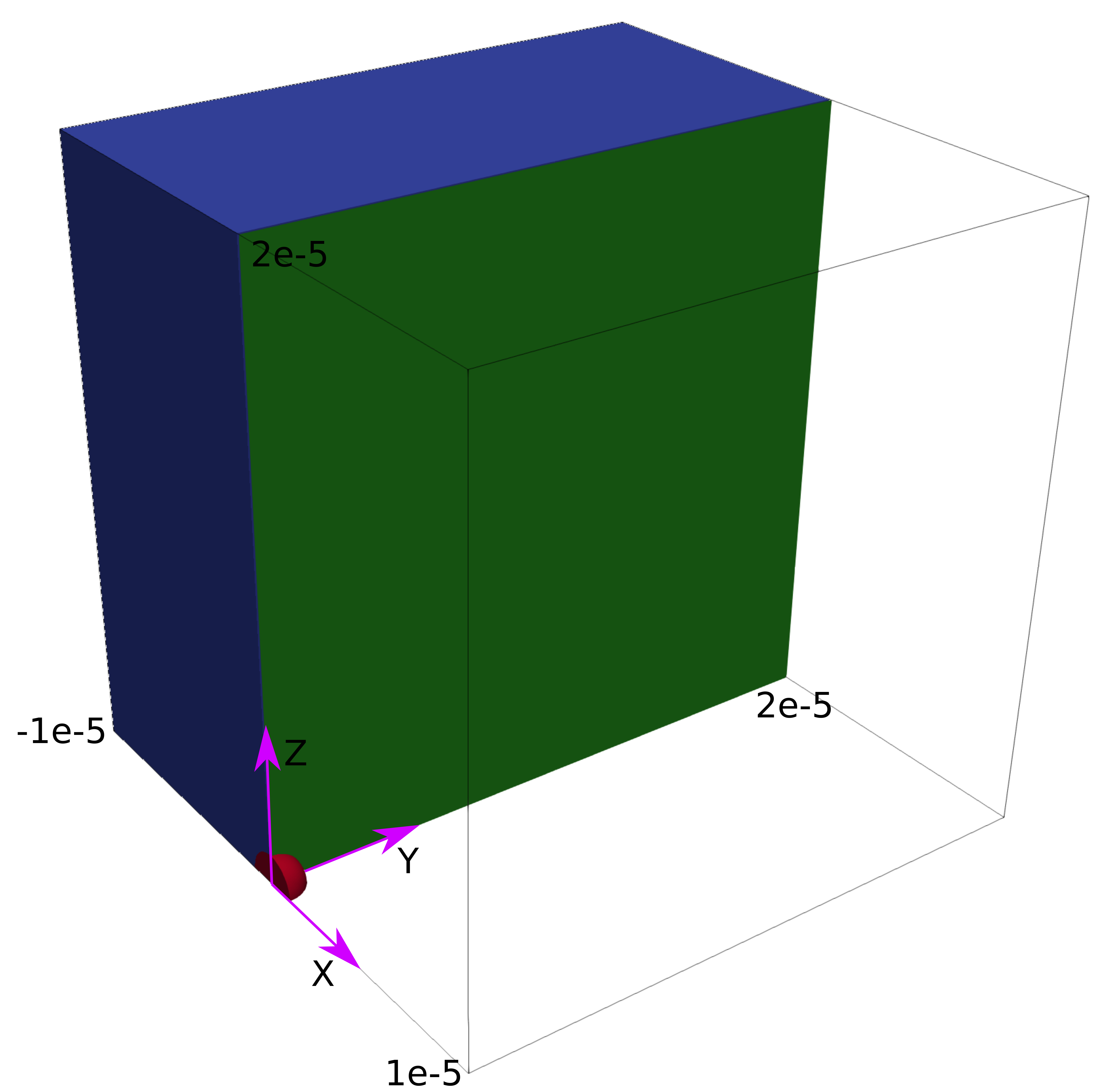}
  \caption{Geometry for the numerical model used for verification. Red
    colour corresponds to the ASR product, blue to the surrounding
    aggregate, green to the pre-defined crack plane. Second half of the
    aggregate is omitted for visualisation of the crack plane. Dimensions
    are given in meters.}
  \label{fig:model_fem}
\end{figure}

The geometry used for the numerical model is shown in
\cref{fig:model_fem}. It comprises one quarter of the full
space. Symmetry over $xz$ and $xy$ planes is established by imposing zero
out-of-plane displacements. The simulated block comprises one quarter of
the spheroidal inclusion and the surrounding aggregate. The equatorial
radius of the spheroid is $1 \ \mu m$, while the dimensions of the full
block are $20 \times 20 \times 20 \ \mu m^3$. Behaviour of the bulk
materials (aggregate and inclusion) is linear elastic.

\begin{figure}
  \centering
  \begin{subfigure}{.5\textwidth}
    \centering \includegraphics[width=\linewidth]{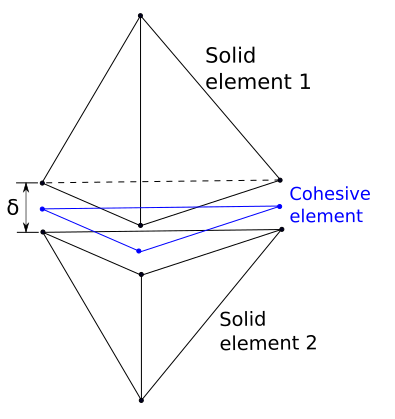}
    \caption{}
    \label{fig:coh_el}
  \end{subfigure}%
  \begin{subfigure}{.5\textwidth}
    \centering \includegraphics[width=\linewidth]{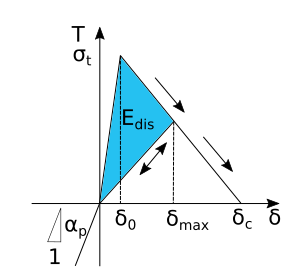}
    \caption{}
    \label{fig:coh_law}
  \end{subfigure}
  \caption{a) A sketch of a first-order 2D cohesive element between 3D
    solid elements. b)Linear cohesive law.}
  \label{fig:coh}
\end{figure}

For verification of the maximum crack radius, the crack plane is added to
the FE model along the $yz$ plane. The discontinuity is modelled by
cohesive elements approach. The latter yields to dissipation of the
fracture energy by interface elements placed between neighbouring solid
elements (see \cref{fig:coh_el}). When two solid elements are being
pulled apart, a cohesive element in-between resists by generating the
inwards traction whose amplitude depends on the opening.  The cohesive
elements represent the non-linear process zone (or cohesive zone) at a
crack tip, as originally proposed by \cite{dugdale_yielding_1960} and
\cite{barenblatt_mathematical_1962}.

Behaviour of cohesive elements is stipulated through the
traction-separation law. For the current study, a bi-linear
traction-separation law, shown in \cref{fig:coh_law}, is employed. For
the cohesive opening, $\delta$, below the elastic limit $\delta_0$, a
cohesive element behaves in a linear elastic manner. After the traction
$T$ reaches the material strength in tension $\sigma_t$, it decreases
linearly, capturing material softening. The latter takes place up to a
moment when opening reaches its critical value $\delta_c$. Then the
element is fully broken. If the loading is removed when the element is in
the softening stage ($\delta < \delta_c$), its behaviour becomes elastic
again within the window $0 \leq \delta \leq \delta_{max}$. Under
compression, cohesive elements behave elastically with a penalty
coefficient $\alpha_p$. Opening of a cohesive element dissipates energy
$E_{dis}$. The maximum value that can be dissipated upon reaching the
critical opening $\delta_c$ equals the fracture energy $\fren$. The
latter could be computed from the fracture toughness $\tough$ as
\begin{equation}
  \label{eq:fract_energy}
  \fren = \frac{\tough^2}{E},
\end{equation}
where $E$ is the Young's modulus of surrounding material. It is assumed that no
friction is present and only normal opening of cohesive elements takes
place.

For a cohesive element model to converge to the LEFM solution, several
requirements have to be fulfilled. The size of the cohesive zone, $l_z$,
has to be smaller than all other model length scales (sizes of the
inclusion, the crack and the domain). The crack size itself has to be
smaller than the domain size. From the numerical side, the cohesive zone
has to be sufficiently resolved and comprise several cohesive
elements. Size of the cohesive zone can be estimated as:
\begin{equation}
  \label{eq:coh_zone}
  l_z = \frac{E \fren}{\sigma_t^2}.
\end{equation}
For an inclusion of $1 \ \mu m$ radius and a domain of $20 \ \mu m$ size,
a $0.2 \ \mu m$ cohesive zone was picked. The mesh along the crack plane
was densified down to $0.03 \ \mu m$, which allowed to have multiple
elements within the cohesive zone. The cohesive elements are inserted
along the pre-defined plane in the beginning of the simulation.

\section{Results}
\label{sec:results}
The theoretical predictions of stresses and crack radii are computed for
different inclusion shapes and sizes. The theory is then validated by
comparing with the numerical model.

\subsection{Stresses in the bulk}
\label{sec:stresses-in-bulk}

To compute the stresses, material properties given in \cref{tab:prop} are
used. Due to unavailability of the mechanical properties of the amorphous
ASR product, those of the crystalline state are taken.
\begin{table}
  \centering
  \begin{tabular}{l c c c}
		
    \textbf{Material properties } & ASR product* &
                                                   Aggregate
                                                   $\dagger$
    & Cohesive material\\
    \hline 
    Young's modulus $E$, [GPa] & $10$  & $60$ & - \\ 
    Poisson's ratio $\nu$, [-] & $0.3$ & $0.3$ & - \\
    Fracture toughness $\tough$, [MPa $\cdot$ m\textsuperscript{1/2}]  &
                                                                         -
                                                 & $2$ & -
    \\
    Fracture energy $\fren$, [J/m$^2$]  & - & - & $66.7$ \\ 
    Tensile strength $\sigma_t$, [MPa] & - & - & $4.5 \cdot 10^3$ \\
    Elastic opening $\delta_0$, [$\eta m$]  & - & - & $1$\\
    Critical opening $\delta_c$, [$\eta m$]  & - & - & $29.6$\\
    
    \hline
  \end{tabular}
  \caption{Material properties of the ASR product, aggregate matrix and
    parameters of the bi-linear cohesive law. * Based on
    micro-indentation tests by \citep{leemann_e-modulus_2013}. $\dagger$
    Taken from \citep{alehossein_strength_2004}.}
  \label{tab:prop}
\end{table}

\begin{figure}
  \centering
  \includegraphics[width=0.8\textwidth]{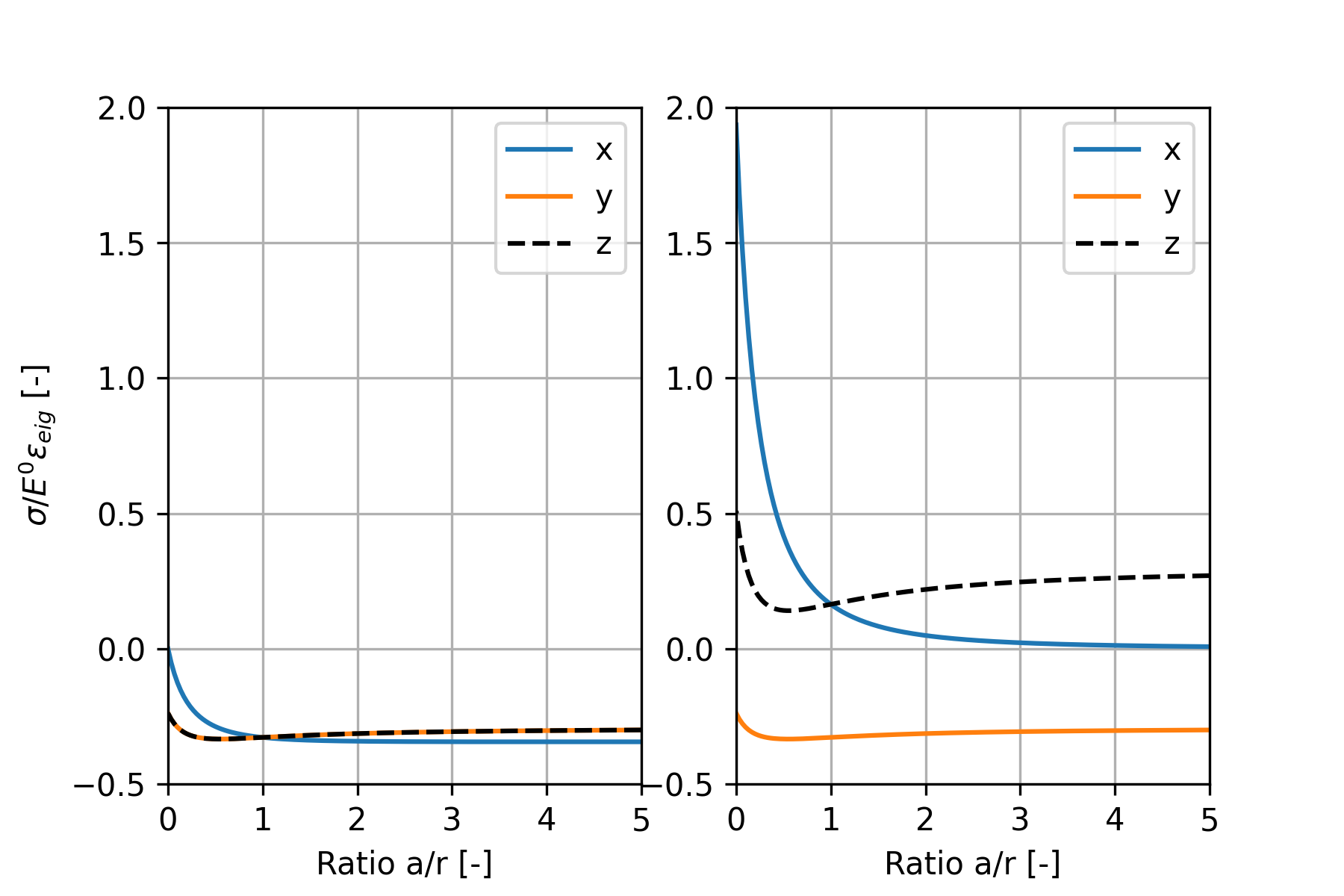}
  \caption{Principle stress values inside (on the left) and outside the
    inclusion (on the right) at the observation point (see
    \cref{fig:ellipsoid}) depending on the direction. Stresses are
    normalised by a product of the matrix's Young's modulus $E^0$ and the
    eigen strain scalar $\varepsilon_{eig}$. The semi-axis $a$ is
    normalised by the equatorial radius $r$.}
  \label{fig:stress_size_depend}
\end{figure}

The solution for the principal stresses inside and outside the inclusion
is given in \cref{fig:stress_size_depend}.
Stresses at the interior points are independent of the point's
position. Stresses in the matrix are plotted at the intersection point
between the $y$-axis and the surface of the ellipsoid (see the
observation point in \cref{fig:ellipsoid}). The semi-axis $a$ is varied
from almost zero to five times the equatorial radius $r$, changing from a
penny to a sphere and finally to a needle. Stresses inside the inclusion
are always compressive, becoming hydrostatic
($\sigma_{xx} = \sigma_{yy} = \sigma_{zz}$) in the spherical
configuration. When approaching the disk limit, stress in the direction
normal to the disk plane tends towards zero. The stress state in the
matrix is different. While the stress in $y$-direction is still
compressive, the other two stresses are either null or tensile. This
creates necessary premises for crack appearance. In the spherical set-up,
tensile stresses in $x$- and $y$-directions are equal, which makes any
plane passing through the $y$-axis suitable for the crack growth. In the
penny-shape limit ($a/r \to 0$), $\sigma_{xx}$ intensifies making
$yz$-plane preferable for a crack to grow. Since we expect the gel
inclusions to have a flat shape, we will limit our analysis to cases when
$a/r$ changes from zero to one.

\begin{figure}
  \centering \includegraphics[width=0.8\textwidth]{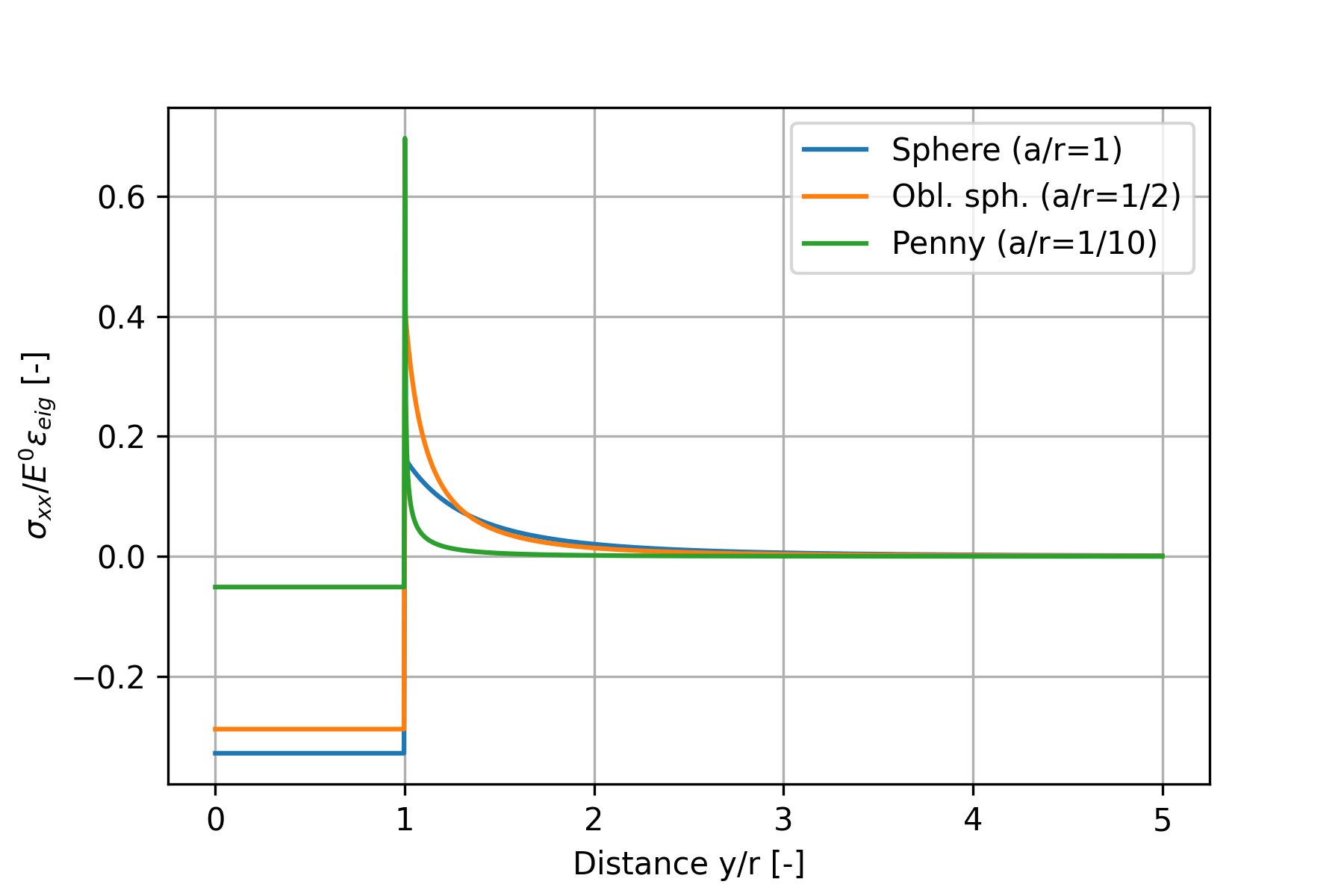}
  \caption{Stress $\sigma_{xx}$ for three different inclusion shapes
    plotted along the $y$-axis. Its value is normalised by a product of
    the matrix's Young's modulus $E^0$ and the eigen strain scalar
    $\varepsilon_{eig}$. The distance $y$ is normalised by the equatorial
    radius $r$.}
  \label{fig:str_ins_outs}
\end{figure}

Stresses along the $y$-axis for three different shapes of the inclusion
are plotted in \cref{fig:str_ins_outs}. Although the disk shape has a
tensile stress at the border much larger than a sphere, it decays much
faster. For the spherical configuration ($a = b = c = r$), the stress
drops with rate $r^3/y^3$, which matches a well-known analytical solution
\citep[p.~417]{timoshenko_theory_1951}.
A particularity of the Eshelby solution is its size independence: both
small and large inclusions under the same expansion value will generate
similar stress and strain fields.

Numerical solution for the stresses along the $y$-axis fairly matches the
analytical predictions. Comparison between two sets of results are
plotted in \cref{fig:stress_comp} of Appendix. In this solution, no crack
is present in the simulation. The only difference between two methods is
observed at the boundary, where FEM cannot represent the stress
singularity for oblate shapes. Refining the mesh improves the
solution. The stress decay further from the inclusion surface is
perfectly resolved.

\subsection{Crack radius}
\label{sec:crack-radius}

The crack is added to the previous problem by means of cohesive
elements. The parameters of the cohesive law are listed in
\cref{tab:prop}. In order to have a small $l_z$ and preserve the fracture
energy, the tensile strength $\sigma_t$ had to be assigned a high value
and the critical opening $\delta_c$ had to be adjusted.

\begin{figure}
  \centering \includegraphics[width=0.8\textwidth]{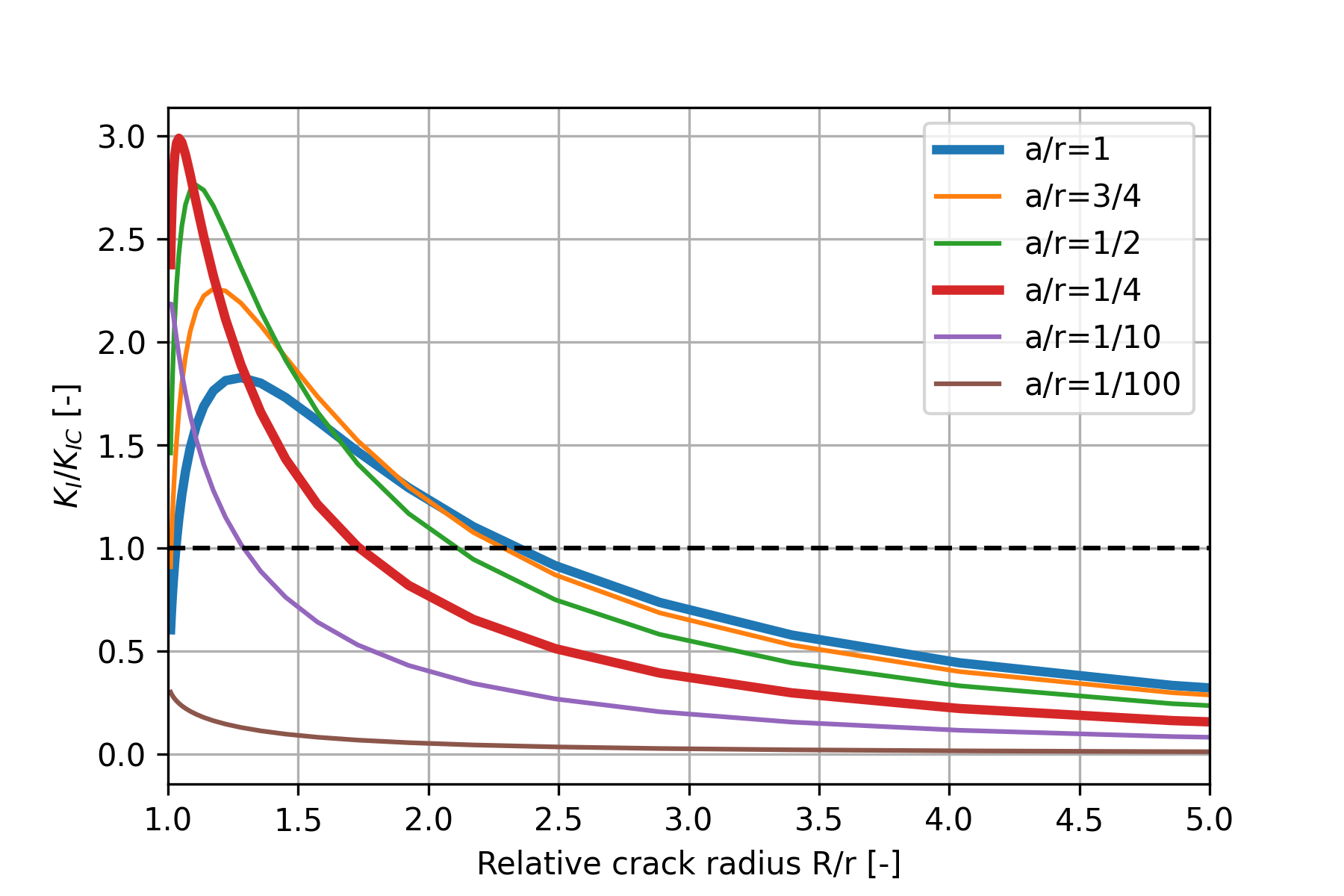}
  \caption{$\sif1/\tough$ depending on the aspect ratio of the spheroid
    $a/r$ and the external crack radius $R/r$ for a spheroid with the
    fixed equatorial radius $r=1 \ \mu m$ under expansion of $100\%$ .}
  \label{fig:sifs}
\end{figure}

Expansion of the inclusion leads to concentration of stresses at the
crack tip. The intensity of this concentration is characterised by the
stress intensity factor. Solutions for $\sif1$ depending on the spheroid
shape and the crack radius are plotted in \cref{fig:sifs}. The equatorial
radius of the inclusion in this plot equals $1$ $\mu$m. $\sif1$
normalised by the fracture toughness $\tough$ serves as an indicator for
crack growth. While values of $\sif1/\tough$ above one suggest that the
crack will extend, strict equality to one indicates the crack radius at
which the growth will cease.

All the curves in \cref{fig:sifs} have unimodal distribution shape. At
the extreme values of R, zero and $\infty$, $\sif1$ tends towards
zero. In physical terms, it means that for a crack to grow, a
pre-existing fissure of a finite length must exist. Otherwise, the
problem transitions from being toughness-controlled to
strength-controlled \citep{bazant_size_1984,ritchie_conflicts_2011}.
Overcoming material strength in the absence of micro-fissures would
require higher expansion values of the ASR pocket. In contrast,
sufficiently large size of a pre-existing crack encircling the inclusion
facilitates its further growth. The curves in \cref{fig:sifs} are
ascending up to the $\sif1$ peak value and later descend to zero. 

The $\sif1$ peak position and its height vary for different inclusion
shapes. The highest $\sif1$ corresponds to an oblate spheroid with the
aspect ratio $a/r = 1/4$ which makes it the most critical case for high
fracture toughness. A sphere ($a/r = 1$) has a lower peak value, but its
maximum crack length is the largest. Consequently, a sphere and a $1/4$
spheroid are the two most critical geometries with regards to the $\sif1$
intensity and the maximum crack length correspondingly.

For the current equatorial radius of a spheroid $b=1 \ \mu$m, any aspect
ratio $1/10 \le a/r \le 1$ leads to the fracture growth. Flattening the
shape of the inclusion would reduce its potential to grow the crack. For
instance, a penny with thickness $a = 0.01 \ \mu$m is not expected to
develop a crack. The maximum crack radius for all possible spheroids with
$1 \ \mu m$ radius is $2.35 \ \mu m$ which is more than twice wider than
the ASR pocket itself.

\begin{figure}
  \centering \includegraphics[width=0.8\textwidth]{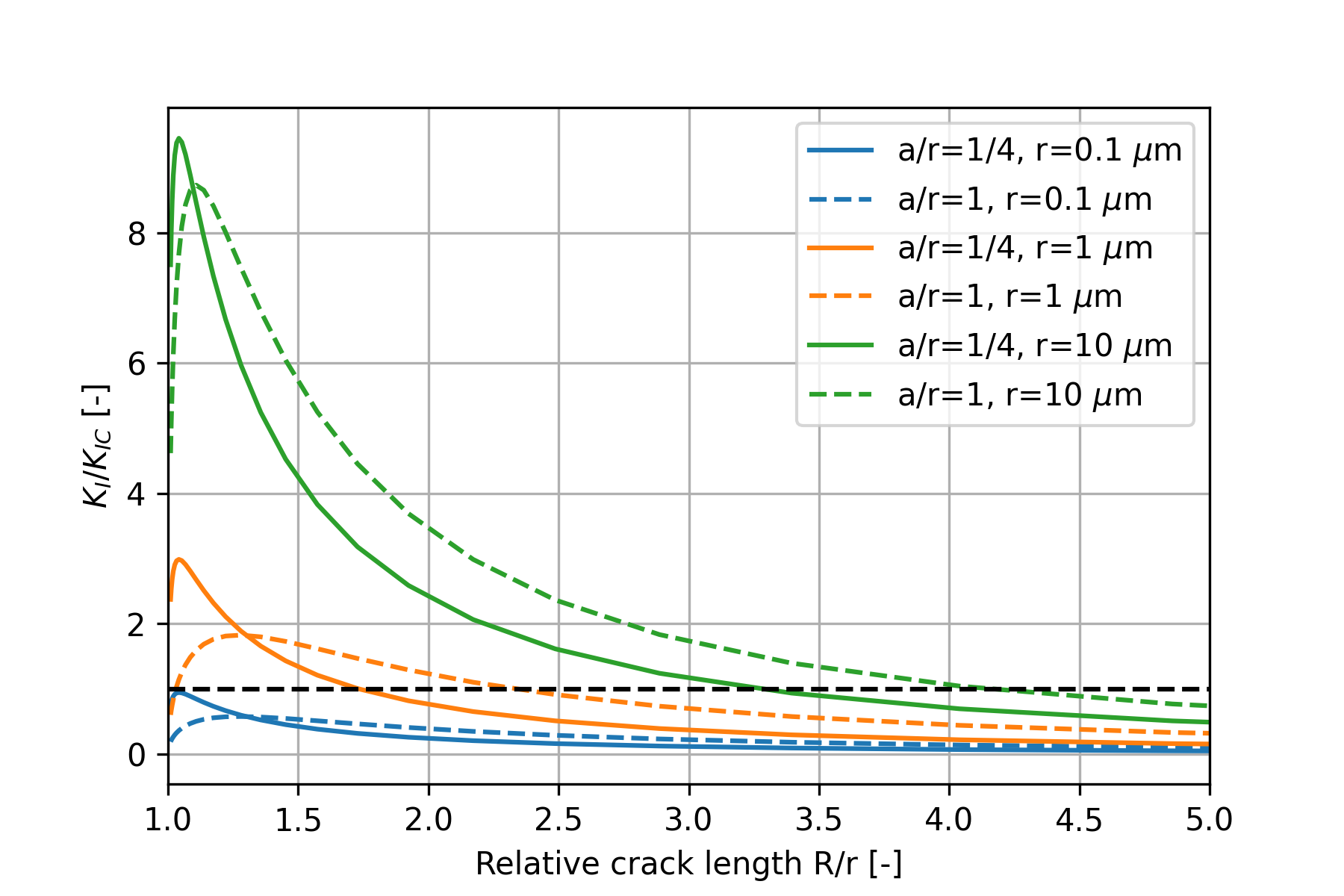}
  \caption{$\sif1/\tough$ for a sphere and $1/4$ spheroid of different radii and
    a constant expansion of $100\%$.}
  \label{fig:sif_size}
\end{figure}
Although the stress solution of Eshelby problem does not depend on the
size of the inclusion, the solution for the stress intensity factor
does. This is demonstrated in \cref{fig:sif_size}. The two most critical
shapes (a sphere and a $1/4$ spheroid) of different radii were tested:
$0.1 \ \mu m$, $1 \ \mu m$, and $10 \ \mu m$. Same expansion value was
applied. One can observe a gradual increase in both the peak SIF value
and the maximum crack length for bigger inclusions. This allows us to
conclude that the size of the ASR pocket has a direct effect on the crack
propagation: larger ASR inclusions have higher chance to damage the
surrounding aggregate and thus cause a longer crack. Another important
observation is that, irrespective of its shape, an inclusion of
$0.1 \ \mu m$ size or smaller does not trigger crack growth. This
observation suggests that there is a critical value of the spheroid
radius below which no crack extension will happen for expansion values of
$100 \%$ and the chosen material properties. This critical radius lays in
the range between $0.1 \ \mu m$ and $1 \ \mu m$.

\begin{figure}
  \centering \includegraphics[width=0.8\textwidth]{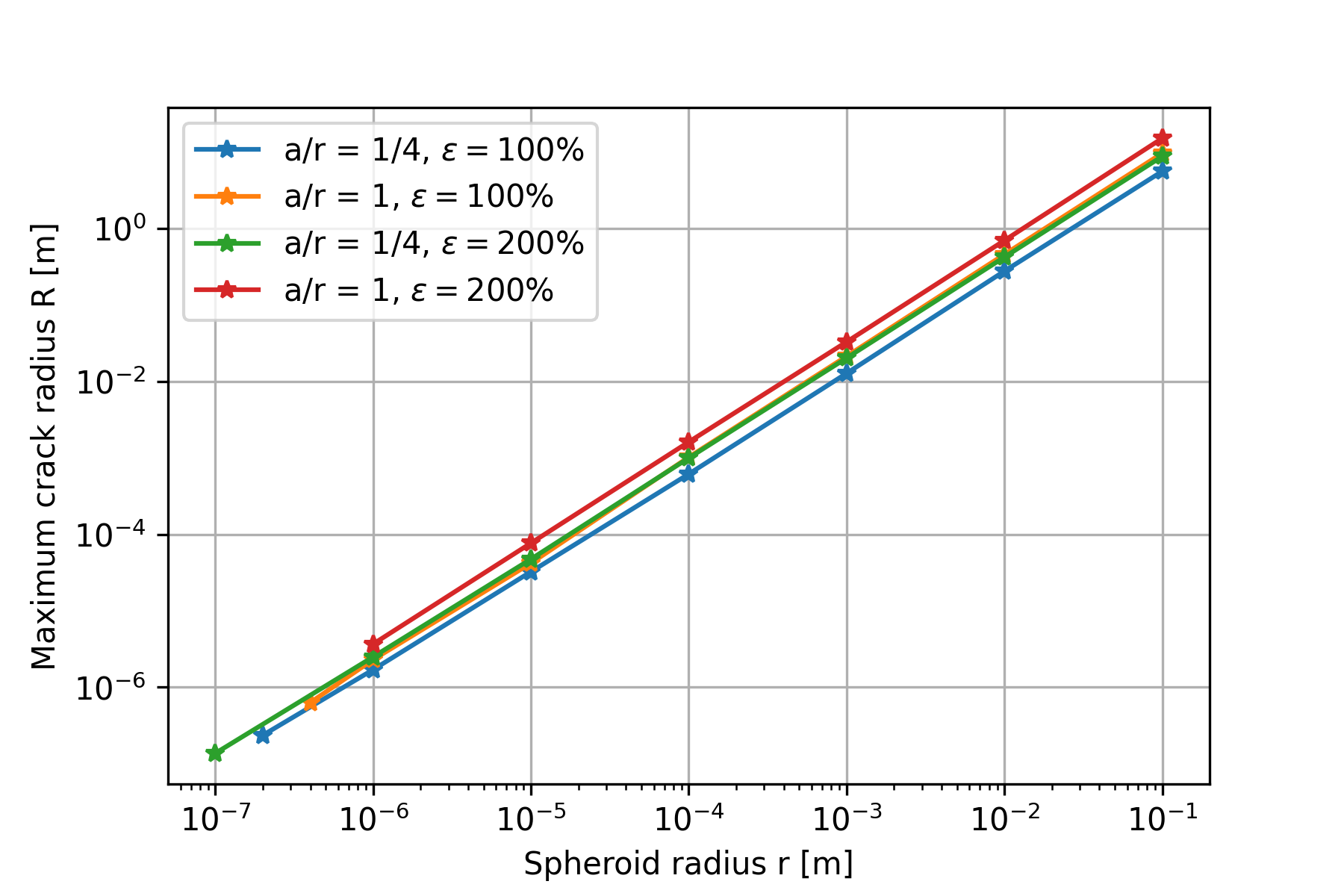}
  \caption{Maximum crack radii for spheroidal inclusions of different
    sizes under different expansion values plotted in log-log scale.}
  \label{fig:max_crack_radii}
\end{figure}
To study how the inclusion size is related to the maximum developed crack
radius, an additional parametric study was performed, results of which
are presented in \cref{fig:max_crack_radii}. Two previously highlighted
shapes (a sphere and a $1/4$ spheroid) were increased in size from
hundreds of nanometers to few centimetres and the corresponding maximum
crack radii were measured. Such a wide range was chosen for purely
illustrative purposes and does not have any physical
implications. Expansion values of $100\%$ and $200\%$ were
applied. Results are plotted in log-log scale. The maximum crack radius
shows power-law dependence on the inclusion size. The power law reveals an
exponent of $4/3$. The
results suggest that the same power-law exponent holds for different
expansion values.

\begin{figure}
  \centering \includegraphics[width=0.7\textwidth]{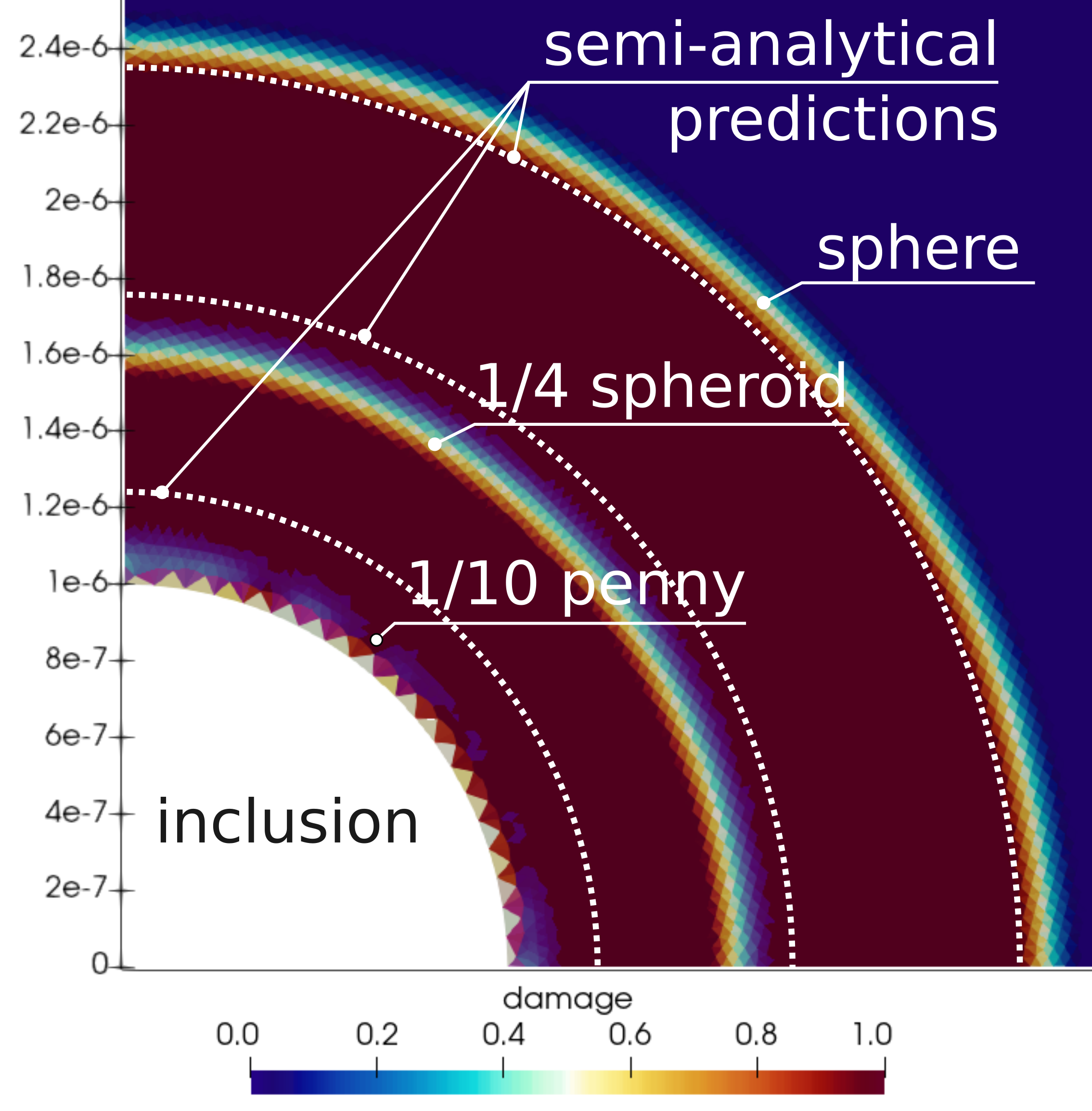}
  \caption{Final values of damage in numerical simulations for three
    spheroids with radius $1 \ \mu m$ and expansion of $100 \%$. Dotted
    lines denote semi-analytical predictions of maximum crack radii
    presented in \cref{fig:sifs}. Dimensions are given in meters.}
  \label{fig:damage_numerical}
\end{figure}

Numerical validation of obtained solution is given in
\cref{fig:damage_numerical}. Here, white dotted lines represent
semi-analytical predictions for three spheroids, while the coloured
background comprises three numerical crack profiles. Final positions of
the numerical cracks can be interpreted through the damage values: one
corresponds to fully damaged cohesive elements, zero is for an intact
material. Rainbow-coloured regions correspond to the cohesive zones at
the external crack radii. The radius of the crack next to a sphere is
about $2.4 \ \mu m$ which is almost equal to the analytically predicted
radius of $2.35 \ \mu m$. For a $1/4$ spheroid, numerically predicted
crack measures $1.6 \ \mu m$ which is $8.5\%$ smaller than the
analytical value of $1.75 \ \mu m$. Finally, a penny has not developed a
full crack and just slightly opened a rim of elements next to it while
its analytical prediction is $1.27 \ \mu m$. The reason for the
underdeveloped crack is the cohesive zone size being equal to the size of
the pre-existing crack. To improve this result, one could reduce the
process zone size even more, which would require to refine the numerical
mesh. Except for this last case, the numerical model fairly reproduces the
semi-analytical predictions.

\section{Conclusions}
\label{sec:conclusions}
In this study, we have estimated the potential of ASR pockets of
different sizes and shapes to cause cracking of surrounding aggregates
due to their volume increase of $100\%$. For this, we have proposed a
semi-analytical model consisting of the expanding ellipsoidal inclusion
and a ring-shaped crack encircling it. The inclusion here represents the
pocket of ASR product at its mature solid state. A prerequisite for such
model is the presence of pre-existing fissures, which are abundant in
rocks. First, the stresses in the matrix are computed by an analytical
solution. Later, the product of computed stresses and the specific weight
function are integrated over the surface of a crack to obtain the stress
intensity factor for mode I. The latter characterises the crack potential
to grow. The maximum distance at which stress intensity factor normalised
by the fracture toughness equals one, renders the maximum crack length.

The resulting solution strongly depends on the inclusion's shape and
size. Flat shapes concentrate higher tensile stresses on their sharp
edges than the bulging ones. At the same time, these stresses decay much
faster further from the inclusion. While the stress intensity factor for
mode I is the highest for an oblate spheroid with an aspect ratio equal
to $1/4$, the longest crack is caused by a spherical shape. It makes
$1/4$ spheroid and a sphere the two most critical shapes, either for
reaching high values of fracture toughness or for causing the longest
crack.

Size of the inclusion plays a crucial role in crack growing
potential. For an inclusion of $1 \ \mu m$ radius under expansion of
$100\%$, all spheroidal shapes with aspect ratio above $1/10$ are likely
to grow a crack around. Flatter shapes would cause either shorter cracks
($R/r \leq 1.25$) or no crack at all. There is a critical inclusion
radius in the range between $0.1 \ \mu m$ and $1 \ \mu m$ below which no
crack growth is expected for the current combination of $\tough$ and
$\varepsilon_{eig}$. Consistently, an increase in the inclusion's size
yields higher values of $\sif1$ and longer cracks.

Further increase in the inclusion size as well as expansion value
suggests a power-law dependence between the radius of a spheroid and the
maximum crack radius. Independent of the spheroidal shape or the
expansion value, the power-law exponent is about $4/3$.

Results of the analytical and semi-analytical predictions were confirmed
by the numerical model. For verification of the stresses in the bulk in
the absence of the crack, a purely elastic finite element model was
assembled. For validating the predicted maximum crack length, this model
was enriched with cohesive elements. Both the elastic stresses and the
crack lengths obtained numerically have a good match with the
semi-analytical results. 

\section*{Acknowledgements}
The Swiss National Science Foundation is acknowledged for financial
support within the Sinergia project ``Alkali-silica reaction in concrete
(ASR)'' through grant CRSII5\_17108.

\newpage
\bibliographystyle{abbrvnat} \bibliography{library}

\begin{thebibliography}{35}
\providecommand{\natexlab}[1]{#1}
\providecommand{\url}[1]{\texttt{#1}}
\expandafter\ifx\csname urlstyle\endcsname\relax
  \providecommand{\doi}[1]{doi: #1}\else
  \providecommand{\doi}{doi: \begingroup \urlstyle{rm}\Url}\fi

\bibitem[noa(2021)]{noauthor_akantu_2021}
Akantu ({Version} 4.0.0), Aug. 2021.
\newblock URL \url{https://akantu.ch/}.

\bibitem[Alehossein and Boland(2004)]{alehossein_strength_2004}
H.~Alehossein and J.~N. Boland.
\newblock Strength, toughness, damage and fatigue of rock.
\newblock In \emph{Structural {Integrity} and {Fracture}: {Proceedings} of the
  {International} {Conference}, {SIF} 2004}, page~8, 2004.

\bibitem[Barenblatt(1962)]{barenblatt_mathematical_1962}
G.~Barenblatt.
\newblock The {Mathematical} {Theory} of {Equilibrium} {Cracks} in {Brittle}
  {Fracture}.
\newblock In \emph{Advances in {Applied} {Mechanics}}, volume~7, pages 55--129.
  Elsevier, 1962.
\newblock ISBN 978-0-12-002007-2.
\newblock \doi{10.1016/S0065-2156(08)70121-2}.
\newblock URL
  \url{https://linkinghub.elsevier.com/retrieve/pii/S0065215608701212}.

\bibitem[Ba{\v z}ant(1984)]{bazant_size_1984}
Z.~P. Ba{\v z}ant.
\newblock Size {Effect} in {Blunt} {Fracture}: {Concrete}, {Rock}, {Metal}.
\newblock \emph{Journal of Engineering Mechanics}, 110\penalty0 (4):\penalty0
  518--535, Apr. 1984.
\newblock ISSN 0733-9399, 1943-7889.
\newblock \doi{10.1061/(ASCE)0733-9399(1984)110:4(518)}.
\newblock URL
  \url{http://ascelibrary.org/doi/10.1061/%28ASCE%290733-9399%281984%29110%3A4%28518%29}.

\bibitem[Bueckner(1970)]{bueckner_novel_1970}
H.~F. Bueckner.
\newblock Novel principle for the computation of stress intensity factors.
\newblock \emph{Zeitschrift fuer Angewandte Mathematik \& Mechanik},
  50\penalty0 (9), Sept. 1970.
\newblock URL \url{https://trid.trb.org/view/3976}.

\bibitem[Cole and Lancucki(1983)]{cole_products_1983}
W.~F. Cole and C.~J. Lancucki.
\newblock Products formed in an aged concrete the occurrence of okenite.
\newblock \emph{Cement and Concrete Research}, 13\penalty0 (5):\penalty0
  611--618, Sept. 1983.
\newblock ISSN 0008-8846.
\newblock \doi{10.1016/0008-8846(83)90049-2}.
\newblock URL
  \url{https://www.sciencedirect.com/science/article/pii/0008884683900492}.

\bibitem[D{\"a}hn et~al.(2016)D{\"a}hn, Arakcheeva, Schaub, Pattison, Chapuis,
  Grolimund, Wieland, and Leemann]{dahn_application_2016}
R.~D{\"a}hn, A.~Arakcheeva, P.~Schaub, P.~Pattison, G.~Chapuis, D.~Grolimund,
  E.~Wieland, and A.~Leemann.
\newblock Application of micro {X}-ray diffraction to investigate the reaction
  products formed by the alkali{\textendash}silica reaction in concrete
  structures.
\newblock \emph{Cement and Concrete Research}, 79:\penalty0 49--56, Jan. 2016.
\newblock ISSN 0008-8846.
\newblock \doi{10.1016/j.cemconres.2015.07.012}.
\newblock URL
  \url{https://www.sciencedirect.com/science/article/pii/S0008884615002094}.

\bibitem[De~Ceukelaire(1991)]{de_ceukelaire_determination_1991}
L.~De~Ceukelaire.
\newblock The determination of the most common crystalline alkali-silica
  reaction product.
\newblock \emph{Materials and Structures}, 24\penalty0 (3):\penalty0 169--171,
  May 1991.
\newblock ISSN 1359-5997, 1871-6873.
\newblock \doi{10.1007/BF02472981}.
\newblock URL \url{http://link.springer.com/10.1007/BF02472981}.

\bibitem[Dugdale(1960)]{dugdale_yielding_1960}
D.~Dugdale.
\newblock Yielding of steel sheets containing slits.
\newblock \emph{Journal of the Mechanics and Physics of Solids}, 8\penalty0
  (2):\penalty0 100--104, May 1960.
\newblock ISSN 00225096.
\newblock \doi{10.1016/0022-5096(60)90013-2}.
\newblock URL
  \url{https://linkinghub.elsevier.com/retrieve/pii/0022509660900132}.

\bibitem[Dunant and Scrivener(2012)]{dunant_effects_2012-1}
C.~F. Dunant and K.~L. Scrivener.
\newblock Effects of uniaxial stress on alkali{\textendash}silica reaction
  induced expansion of concrete.
\newblock \emph{Cement and Concrete Research}, 42\penalty0 (3):\penalty0
  567--576, Mar. 2012.
\newblock ISSN 00088846.
\newblock \doi{10.1016/j.cemconres.2011.12.004}.
\newblock URL
  \url{https://linkinghub.elsevier.com/retrieve/pii/S0008884611003267}.

\bibitem[Eshelby(1957)]{eshelby_determination_1957}
J.~D. Eshelby.
\newblock The determination of the elastic field of an ellipsoidal inclusion,
  and related problems.
\newblock \emph{Proc. R. Soc. Lond. A}, 241\penalty0 (1226):\penalty0 376--396,
  Aug. 1957.
\newblock ISSN 0080-4630, 2053-9169.
\newblock \doi{10.1098/rspa.1957.0133}.
\newblock URL
  \url{https://royalsocietypublishing.org/doi/10.1098/rspa.1957.0133}.

\bibitem[Eshelby(1959)]{eshelby_elastic_1959}
J.~D. Eshelby.
\newblock The elastic field outside an ellipsoidal inclusion.
\newblock \emph{Proc. R. Soc. Lond. A}, 252\penalty0 (1271):\penalty0 561--569,
  Oct. 1959.
\newblock ISSN 0080-4630, 2053-9169.
\newblock \doi{10.1098/rspa.1959.0173}.
\newblock URL
  \url{https://royalsocietypublishing.org/doi/10.1098/rspa.1959.0173}.

\bibitem[Fett and Rizzi(2007)]{fett_weight_2007}
T.~Fett and G.~Rizzi.
\newblock Weight {Functions} and {Stress} {Intensity} {Factors} for
  {Ring}-shaped {Cracks}.
\newblock page~48, 2007.

\bibitem[Geng et~al.(2020)Geng, Shi, Leemann, Glazyrin, Kleppe, Daisenberger,
  Churakov, Lothenbach, Wieland, and D{\"a}hn]{geng_mechanical_2020}
G.~Geng, Z.~Shi, A.~Leemann, K.~Glazyrin, A.~Kleppe, D.~Daisenberger,
  S.~Churakov, B.~Lothenbach, E.~Wieland, and R.~D{\"a}hn.
\newblock Mechanical behavior and phase change of alkali-silica reaction
  products under hydrostatic compression.
\newblock \emph{Acta Crystallogr B Struct Sci Cryst Eng Mater}, 76\penalty0
  (4):\penalty0 674--682, Aug. 2020.
\newblock ISSN 2052-5206.
\newblock \doi{10.1107/S205252062000846X}.
\newblock URL \url{https://scripts.iucr.org/cgi-bin/paper?S205252062000846X}.

\bibitem[Griffith(1921)]{griffith_phenomena_1921}
A.~A. Griffith.
\newblock The phenomena of rupture and flow in solids.
\newblock \emph{Phil. Trans. R. Soc. Lond. A}, 221\penalty0 (582-593):\penalty0
  163--198, Jan. 1921.
\newblock ISSN 0264-3952, 2053-9258.
\newblock \doi{10.1098/rsta.1921.0006}.
\newblock URL
  \url{https://royalsocietypublishing.org/doi/10.1098/rsta.1921.0006}.

\bibitem[Healy(2009)]{healy_elastic_2009}
D.~Healy.
\newblock Elastic field in {3D} due to a spheroidal
  inclusion{\textemdash}{MATLAB}{\texttrademark} code for {Eshelby}'s solution.
\newblock \emph{Computers \& Geosciences}, 35\penalty0 (10):\penalty0
  2170--2173, Oct. 2009.
\newblock ISSN 00983004.
\newblock \doi{10.1016/j.cageo.2008.11.012}.
\newblock URL
  \url{https://linkinghub.elsevier.com/retrieve/pii/S0098300409001356}.

\bibitem[Iskhakov et~al.(2019)Iskhakov, Timothy, and
  Meschke]{iskhakov_expansion_2019}
T.~Iskhakov, J.~J. Timothy, and G.~Meschke.
\newblock Expansion and deterioration of concrete due to {ASR}:
  {Micromechanical} modeling and analysis.
\newblock \emph{Cement and Concrete Research}, 115:\penalty0 507--518, Jan.
  2019.
\newblock ISSN 00088846.
\newblock \doi{10.1016/j.cemconres.2018.08.001}.
\newblock URL
  \url{https://linkinghub.elsevier.com/retrieve/pii/S0008884618301509}.

\bibitem[Ju and Sun(1999)]{ju_novel_1999}
J.~W. Ju and L.~Z. Sun.
\newblock A {Novel} {Formulation} for the {Exterior}-{Point} {Eshelby}'s
  {Tensor} of an {Ellipsoidal} {Inclusion}.
\newblock \emph{Journal of Applied Mechanics}, 66\penalty0 (2):\penalty0
  570--574, June 1999.
\newblock ISSN 0021-8936, 1528-9036.
\newblock \doi{10.1115/1.2791090}.
\newblock URL
  \url{https://asmedigitalcollection.asme.org/appliedmechanics/article/66/2/570/445956/A-Novel-Formulation-for-the-ExteriorPoint-Eshelbys}.

\bibitem[Ju and Sun(2001)]{ju_effective_2001}
J.~W. Ju and L.~Z. Sun.
\newblock Effective elastoplastic behavior of metal matrix composites
  containing randomly located aligned spheroidal inhomogeneities. {Part} {I}:
  micromechanics-based formulation.
\newblock \emph{International Journal of Solids and Structures}, page~19, 2001.

\bibitem[Larive(1997)]{larive_apports_1997}
C.~Larive.
\newblock \emph{Apports combin{\'e}s de l'exp{\'e}rimentation et de la
  mod{\'e}lisation {\`a} la compr{\'e}hension de l'alcali-r{\'e}action et de
  ses effets m{\'e}caniques}.
\newblock PhD thesis, l'{\'E}cole Nationale des Ponts et Chauss{\'e}es, 1997.

\bibitem[Leemann and Lura(2013)]{leemann_e-modulus_2013}
A.~Leemann and P.~Lura.
\newblock E-modulus of the alkali{\textendash}silica-reaction product
  determined by micro-indentation.
\newblock \emph{Construction and Building Materials}, 44:\penalty0 221--227,
  July 2013.
\newblock ISSN 09500618.
\newblock \doi{10.1016/j.conbuildmat.2013.03.018}.
\newblock URL
  \url{https://linkinghub.elsevier.com/retrieve/pii/S0950061813002237}.

\bibitem[Leemann and M{\"u}nch(2019)]{leemann_addition_2019}
A.~Leemann and B.~M{\"u}nch.
\newblock The addition of caesium to concrete with alkali-silica reaction:
  {Implications} on product identification and recognition of the reaction
  sequence.
\newblock \emph{Cement and Concrete Research}, 120:\penalty0 27--35, June 2019.
\newblock ISSN 00088846.
\newblock \doi{10.1016/j.cemconres.2019.03.016}.
\newblock URL
  \url{https://linkinghub.elsevier.com/retrieve/pii/S0008884619300730}.

\bibitem[Leemann et~al.(2016)Leemann, Katayama, Fernandes, and
  Broekmans]{leemann_types_2016}
A.~Leemann, T.~Katayama, I.~Fernandes, and M.~A. T.~M. Broekmans.
\newblock Types of alkali{\textendash}aggregate reactions and the products
  formed.
\newblock \emph{Proceedings of the Institution of Civil Engineers -
  Construction Materials}, 169\penalty0 (3):\penalty0 128--135, June 2016.
\newblock ISSN 1747-650X, 1747-6518.
\newblock \doi{10.1680/jcoma.15.00059}.
\newblock URL
  \url{http://www.icevirtuallibrary.com/doi/10.1680/jcoma.15.00059}.

\bibitem[Leemann et~al.(2020)Leemann, Shi, and Lindg{\r
  a}rd]{leemann_characterization_2020}
A.~Leemann, Z.~Shi, and J.~Lindg{\r a}rd.
\newblock Characterization of amorphous and crystalline {ASR} products formed
  in concrete aggregates.
\newblock \emph{Cement and Concrete Research}, 137:\penalty0 106190, Nov. 2020.
\newblock ISSN 00088846.
\newblock \doi{10.1016/j.cemconres.2020.106190}.
\newblock URL
  \url{https://linkinghub.elsevier.com/retrieve/pii/S0008884620304324}.

\bibitem[Multon and Toutlemonde(2006)]{multon_effect_2006}
S.~Multon and F.~Toutlemonde.
\newblock Effect of applied stresses on alkali{\textendash}silica
  reaction-induced expansions.
\newblock \emph{Cement and Concrete Research}, 36\penalty0 (5):\penalty0
  912--920, May 2006.
\newblock ISSN 00088846.
\newblock \doi{10.1016/j.cemconres.2005.11.012}.
\newblock URL
  \url{https://linkinghub.elsevier.com/retrieve/pii/S0008884605002838}.

\bibitem[Mura(1987)]{mura_micromechanics_1987}
T.~Mura.
\newblock \emph{Micromechanics of defects in solids}.
\newblock 1987.
\newblock ISBN: 9789400934894 OCLC: 1158181484.

\bibitem[Petroski and Achenbach(1978)]{petroski_computation_1978}
H.~Petroski and J.~Achenbach.
\newblock Computation of the weight function from a stress intensity factor.
\newblock \emph{Engineering Fracture Mechanics}, 10\penalty0 (2):\penalty0
  257--266, Jan. 1978.
\newblock ISSN 00137944.
\newblock \doi{10.1016/0013-7944(78)90009-7}.
\newblock URL
  \url{https://linkinghub.elsevier.com/retrieve/pii/0013794478900097}.

\bibitem[Rice(1972)]{rice_remarks_1972}
J.~R. Rice.
\newblock Some remarks on elastic crack-tip stress fields.
\newblock \emph{International Journal of Solids and Structures}, 8\penalty0
  (6):\penalty0 751--758, June 1972.
\newblock ISSN 00207683.
\newblock \doi{10.1016/0020-7683(72)90040-6}.
\newblock URL
  \url{https://linkinghub.elsevier.com/retrieve/pii/0020768372900406}.

\bibitem[Richart and Molinari(2015)]{richart_implementation_2015}
N.~Richart and J.~Molinari.
\newblock Implementation of a parallel finite-element library: {Test} case on a
  non-local continuum damage model.
\newblock \emph{Finite Elements in Analysis and Design}, 100:\penalty0 41--46,
  Aug. 2015.
\newblock ISSN 0168874X.
\newblock \doi{10.1016/j.finel.2015.02.003}.
\newblock URL
  \url{https://linkinghub.elsevier.com/retrieve/pii/S0168874X15000153}.

\bibitem[Ritchie(2011)]{ritchie_conflicts_2011}
R.~O. Ritchie.
\newblock The conflicts between strength and toughness.
\newblock \emph{Nature Mater}, 10\penalty0 (11):\penalty0 817--822, Nov. 2011.
\newblock ISSN 1476-4660.
\newblock \doi{10.1038/nmat3115}.
\newblock URL \url{https://www.nature.com/articles/nmat3115}.
\newblock Bandiera\_abtest: a Cg\_type: Nature Research Journals Number: 11
  Primary\_atype: Reviews Publisher: Nature Publishing Group Subject\_term:
  Mechanical properties Subject\_term\_id: mechanical-properties.

\bibitem[Shi et~al.(2019)Shi, Geng, Leemann, and
  Lothenbach]{shi_synthesis_2019}
Z.~Shi, G.~Geng, A.~Leemann, and B.~Lothenbach.
\newblock Synthesis, characterization, and water uptake property of
  alkali-silica reaction products.
\newblock \emph{Cement and Concrete Research}, 121:\penalty0 58--71, July 2019.
\newblock ISSN 00088846.
\newblock \doi{10.1016/j.cemconres.2019.04.009}.
\newblock URL
  \url{https://linkinghub.elsevier.com/retrieve/pii/S0008884619301139}.

\bibitem[Shi et~al.(2020)Shi, Park, Lothenbach, and
  Leemann]{shi_formation_2020}
Z.~Shi, S.~Park, B.~Lothenbach, and A.~Leemann.
\newblock Formation of shlykovite and {ASR}-{P1} in concrete under accelerated
  alkali-silica reaction at 60 and 80~{\textdegree}{C}.
\newblock \emph{Cement and Concrete Research}, 137:\penalty0 106213, Nov. 2020.
\newblock ISSN 0008-8846.
\newblock \doi{10.1016/j.cemconres.2020.106213}.
\newblock URL
  \url{https://www.sciencedirect.com/science/article/pii/S0008884620304622}.

\bibitem[Swamy(2003)]{swamy_alkali-silica_2003}
R.~N. Swamy.
\newblock \emph{Alkali-silica reaction in concrete}.
\newblock Blackie and Son ; Taylor and Francis e-Library, Glasgow, England; New
  York, New York, 2003.
\newblock ISBN 978-0-203-20033-9.
\newblock URL \url{https://www.taylorfrancis.com/books/0203036638}.
\newblock OCLC: 1132080496.

\bibitem[Tada et~al.(2000)Tada, Paris, and Irwin]{tada_stress_2000}
H.~Tada, P.~C. Paris, and G.~R. Irwin.
\newblock \emph{The stress analysis of cracks handbook}.
\newblock ASME Press, New York, 3rd ed edition, 2000.
\newblock ISBN 978-0-7918-0153-6.

\bibitem[Timoshenko and Goodier(1951)]{timoshenko_theory_1951}
S.~Timoshenko and J.~N. Goodier.
\newblock \emph{Theory of {Elasticity}}.
\newblock McGraw-Hill, 3 edition, 1951.
\newblock ISBN 978-0-07-085805-3.

\end{thebibliography}

\newpage 
\appendix

\section{Supplementary plots}
\label{sec:supplementary-plots}

\begin{figure}
  \centering \includegraphics[width=0.8\textwidth]{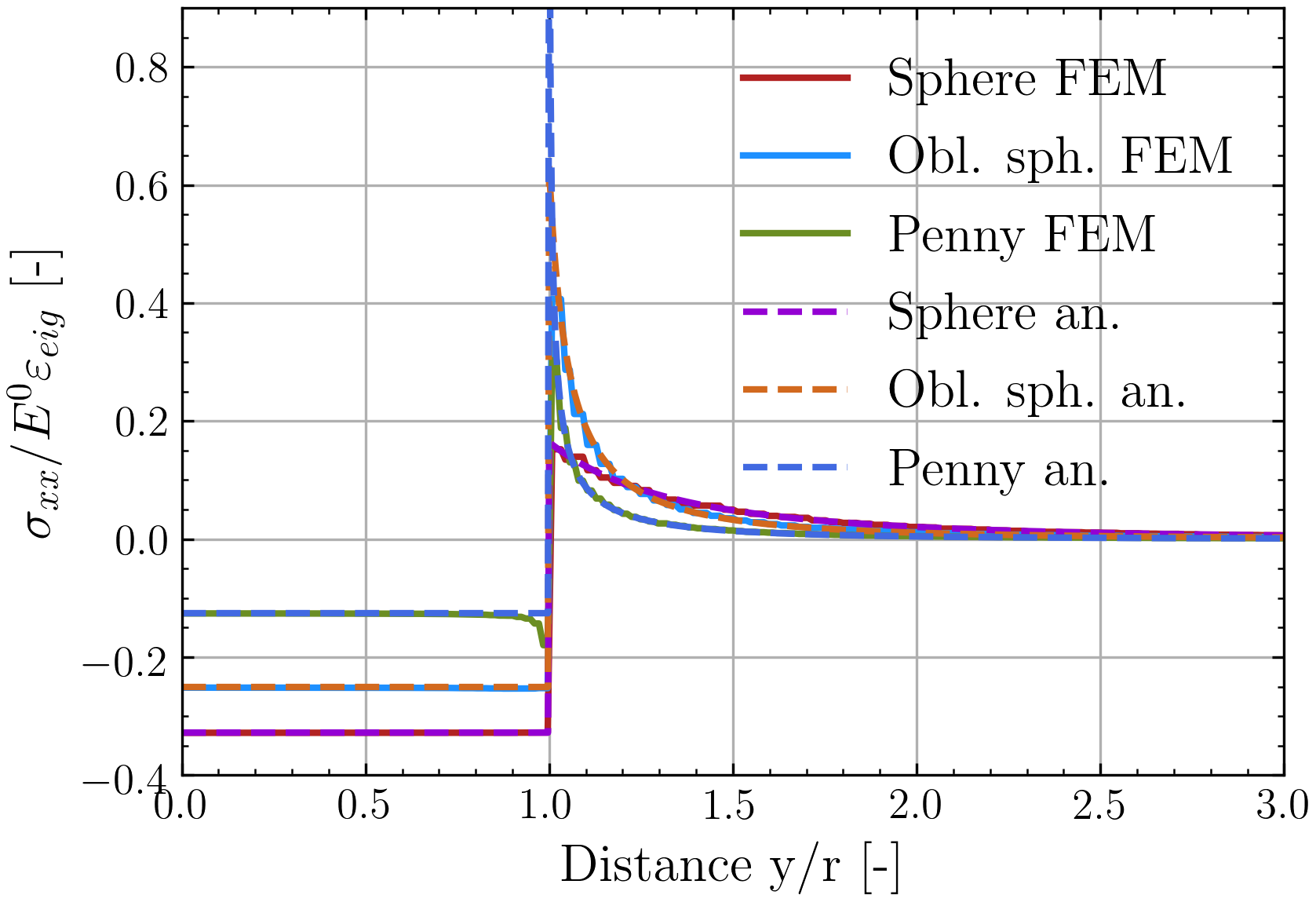}
  \caption{Comparison of stress $\sigma_{xx}$ along $y$-axis from
    analytical model and FE simulation}
  \label{fig:stress_comp}
\end{figure}

\end{document}